\documentclass{iau} 
\usepackage{graphicx}

\title[Quasar detection] 
{Detection of quasars in the time domain}

\author[Graham, Djorgovski, Stern, Drake, \& Mahabal]   
{Matthew J. Graham$^{1,2}$, S. G. Djorgovski$^1$, Daniel J. Stern$^3$, Andrew Drake$^1$ 
\and Ashish Mahabal$^1$}

\affiliation{$^1$California Institute of Technology, \\ Pasadena CA, USA \\ email: {\tt mjg, george, ajd, aam@caltech.edu} \\[\affilskip]
$^2$National Optical Astronomy Observatory, Tucson AZ, USA \\ [\affilskip]
$^3$JPL, Pasadena CA, USA \\ 
email: {\tt daniel.k.stern@jpl.nasa.gov}}

\pubyear{2017}
\volume{325}  
\jname{Astroinformatics}
\editors{M. Brescia, S. G. Djorgovski, E. Feigelson, G. Longo \& S. Cavuoti, eds.}
\begin{document}

\maketitle

\begin{abstract}
The time domain is the emerging forefront of astronomical research with new facilities and instruments providing unprecedented amounts of data on the temporal behavior of astrophysical populations.
Dealing with the size and complexity of this requires new techniques and methodologies. Quasars are an ideal work set for developing and applying these: they vary in a detectable but not easily quantifiable manner whose physical origins are poorly understood. In this paper, we will review how quasars are identified by their variability and how these techniques can be improved, what physical insights into their variability can be gained from studying extreme examples of variability, and what approaches can be taken to increase the number of quasars known. These will demonstrate how astroinformatics is essential to discovering and understanding this important population.

\end{abstract}

\firstsection 
\section{Introduction}

Quasars are a key population for investigating and understanding many astrophysical problems, ranging from theories of galaxy formation and evolution and studies of large-scale structure to the physics of accretion and high energy phenomena. They are studied across the full range of the electromagnetic spectrum as well as being sources of the highest energy cosmic rays, neutrinos, and potentially nanofrequency gravitational waves. Traditionally, they have also been associated with survey astronomy, requiring large sky coverage to define sizable samples (relative to other astronomical populations): first through surveys of the strongest radio sources in the 1960s and 1970s, then surveys of objects with ultraviolet excess in the 1980s and 1990s and, finally, multicolor selection in the (post-)SDSS era. Now time domain surveys are generating new quasar selection methods.


Quasars have long been known to be variable sources; indeed, \cite[Matthews \& Sandage (1963)]{3c48} noted that the most striking feature in optical photometry of 3C 48, one of the first identified quasars, was that the optical radiation varied. Their variability -- photometric, color and spectral -- is well characterized: it is aperiodic in nature, most rapid at the highest energies,  and (anti-)correlated at optical/UV wavelengths with various physical parameters, such as: time lag, rest-frame wavelength, luminosity, radio and emission line properties, the Eddington ratio and estimated black hole mass, e.g., \cite[Ulrich et al. 1997]{ulrich}, \cite[Bauer et al. 2009]{bauer09}, \cite[MacLeod et al. 2010]{macleod10}, \cite[Meusinger et al. 2011]{meusinger11}, \cite[Schmidt et al. 2012]{schmidt12}, \cite[Kelly et al. 2013]{kelly13}. However, the physical mechanisms underlying the (optical/UV) variability remain unclear. It could arise from instabilities in the accretion disk (\cite[Kawaguchi et al. 1998]{kawaguchi}), supernovae (\cite[Aretxaga et al. 1997]{aretxaga}), microlensing (\cite[Hawkins 1993, 2010]{hawkins93, hawkins10}), stellar collisions (\cite[Torricelli-Ciamponi et al. 2000]{torricelli00}), thermal fluctuations from magnetic field turbulence (\cite[Kelly et al. 2009]{kelly09}) or more general (Poisson) processes (\cite[Cid Fernandes et al. 1997]{cidfernandes}). Many studies of quasar variability, though, have been based on small sample sizes with high sampling rates (detailed time series of a few objects) or large samples with (very) sparse sampling, typically just a few epochs.

Quasars and their temporal behavior can thus be seen as a poster child for the burgeoning field of astroinformatics. They require large volumes of heterogeneous archival data and can have a strong real time (transient) aspect to it, particularly at higher energies. Statistically, there are issues of heteroskedasticity, irregular sampling, gappyness, and censored values to deal with. The major science questions also relate to unsupervised and supervised learning through characterization and classification. In this paper, we will review how machine learning and advanced statistical techniques are helping us find and understand these sources and, in particular, the extreme aspects of their variability.

\section{The Catalina Real-time Transient Survey}

Although there are now a number of surveys with sufficient sky and/or temporal coverage to study the ensemble variability properties of quasars, the Catalina Real-time Transient Survey (CRTS, \cite[Drake et al. 2009]{drake09}) enables studies of hundreds of thousands of known (spectroscopically confirmed) quasars with enough resolution to isolate and characterize individual objects as well. It is the largest open (publicly accessible) time domain survey currently operating, leveraging the Catalina Sky Survey data streams from three telescopes --  the 0.7m Catalina Sky Survey (CSS) Schmidt and 1.5m Mount Lemmon Survey (MLS) telescopes in Arizona and  the 0.5m Siding Springs Survey (SSS) Schmidt in Australia -- used in a search for Near-Earth Objects, operated by Lunar and Planetary Laboratory at University of Arizona. 
CRTS covers up to $\sim$2500 deg$^2$ per night, with 4 exposures per visit, separated by 10 min, over 21 nights per lunation, with a total (archival) coverage of $\sim$33,000 deg$^2$ between -75$^\circ <$ Dec $< 70^\circ$ (except for within $\sim$10$^\circ$ -- 15$^\circ$ of the Galactic plane) to a depth of V $\sim$ 19 to 21.5. New cameras in Fall 2016 with larger fields-of-view will increase the nightly sky coverage. 

All data are automatically processed in real-time, and optical transients are immediately distributed using a variety of electronic mechanisms\footnote{http://www.skyalert.org}. The data are broadly calibrated to Johnson $V$ (see \cite[Drake et al. 2013]{drake13} for details) and the full CRTS data set\footnote{http://nesssi.cacr.caltech.edu/DataRelease} contains time series for approximately 500 million sources with a median of $\sim$320 observations per time series over an 11-year baseline. It should be noted that this is roughly the same single filter coverage as LSST will have which makes CRTS an excellent testbed for LSST analyses. In fact, in terms of number of sources and data volume, CRTS is a Moore's law scaled version of LSST. The CRTS archive has light curves for 400,000 spectroscopically confirmed quasars, which represents at least a fivefold improvement over SDSS Stripe 82 data in terms of temporal sampling: this had so far been the definitive data set for quasar variability investigations.

\section{Statistical descriptors}

A number of variability-based features have been employed in the literature for identifying quasars, such as scale and morphological measures, but the two most commonly used in recent analyses are the structure function, e.g., Schmidt et al. (2010), and modeling the variability of the source as a damped random walk (DRW) process, e.g., \cite[Kelly et al. (2009)]{kelly09}. However, these may be too simplistic and algorithms with more statistical power should be considered.

\subsection{Autoregressive models}
Despite their wide use, \cite[Kozlowski (2016a,b)]{kozlowski16a, kozlowski16b} has recently shown that CAR(1) models are statistically degenerate -- they lack discriminatory power and can even provide a better fit to time series generated by non-stochastic processes than the original processes themselves. The values of the characterizing parameters commonly found via MCMC or optimization techniques are also not truly optimal but dependent on the length of the time series (which need to be typically ten times longer than any characteristic timescales). The CAR(1) model is, however, the simplest of the continuous time autoregressive moving average (CARMA) models and higher order models may provide a better description (\cite[Kelly et al. 2015; Kasliwal et al 2016]{kelly15, kasliwal16}): a zero-mean CARMA(p,q) $(p > q)$ process $y(t)$ is defined to be the solution to the stochastic differential equation (CAR(1) = CARMA(1, 0):

\begin{eqnarray*}
\frac{d^py(t)}{dt^p} + \alpha_{p-1} \frac{d^{p-1}y(t)}{dt^{p-1}} + \ldots + \alpha_0y(t) = \\
\beta_q \frac{d^q\epsilon(t)}{dt^q} + \beta_{q-1} \frac{d^{q-1}\epsilon(t)}{dt^{q-1}} + \ldots + \epsilon(t)
\end{eqnarray*}

\noindent
where $\epsilon(t)$ is a continuous time white noise process with zero mean and variance $\sigma^2$.

\begin{figure}[b]
\begin{center}
\includegraphics[width=5.3in]{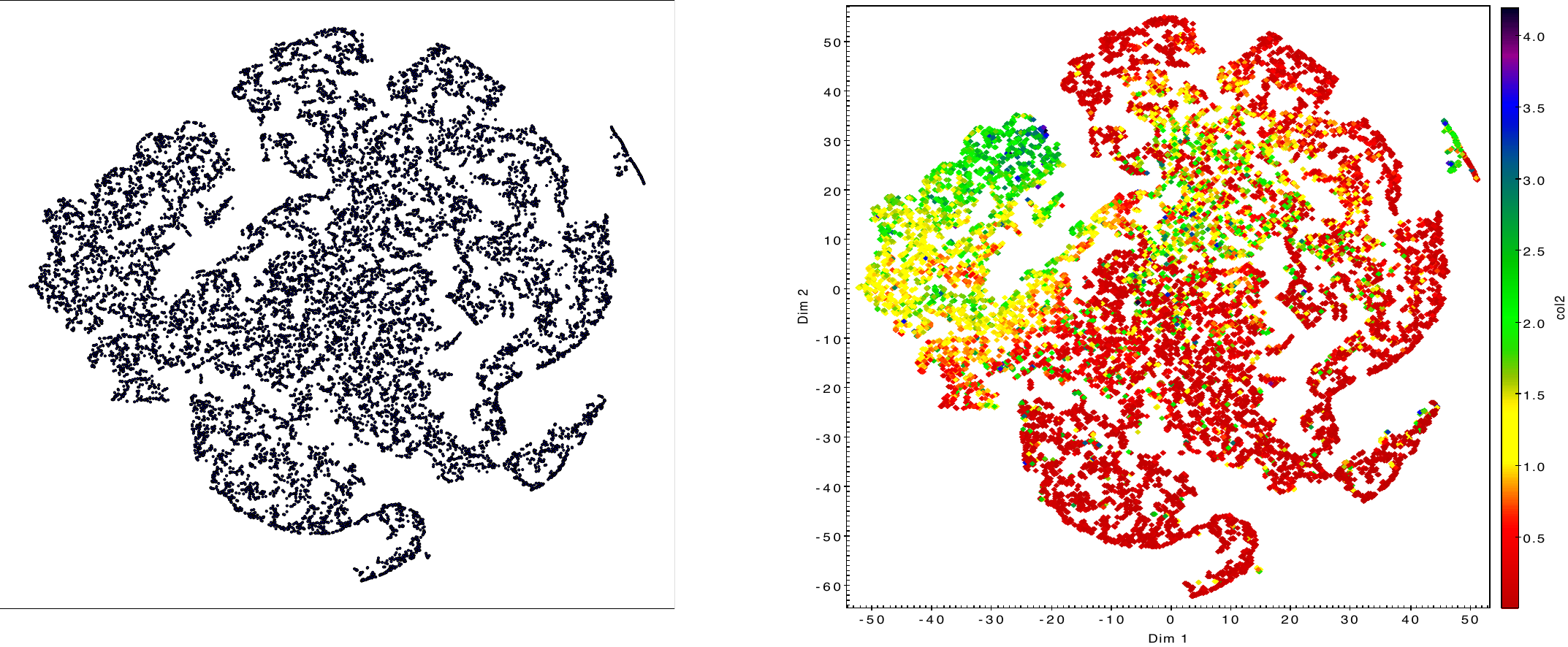} 
\caption{A base TSNE plot (left) of the restframe CARMA(3,2) parameters for 16,498 AGN and with redshift labelled (right).}
\label{fig2}
\end{center}
\end{figure}

CARMA models produce a higher dimensional characterization of a time series through the set of $p + q +1$ parameters $(\alpha_{p-1},...,\alpha_0,\beta_q,...,\beta_1, \sigma^2)$ and appropriate techniques are required to look for relationships between these. T-distributed stochastic neighbor embedding (TSNE, \cite[van der Maarten \& Hinton (2008)]{maarten08}) is an unsupervised dimensional reduction algorithm that preserves topological information. Fig.~\ref{fig2} shows 2-D TSNE plots for the restframe parameters of CARMA(3,2) models for a set of 16,498 CRTS quasars. The base (unlabelled) plot shows structure within the distribution of these parameters and there is clearly localization when the points are labelled with redshift. This may, however, just be a reflection of the timescale coverage of the light curves rather than anything more physical but it certainly warrants further investigation.

\subsection{Slepian wavelet variance}

Wavelets afford a localized time and frequency analysis and a time series can be decomposed by applying a set of wavelet filters:

\[ W_{j,t} = \sum_{l = 0}^{L_j - 1} h_{jl} X_{t-l} ;\,\, t = 0, \pm 1 \ldots; j = 1,2, \ldots; L \ge 2d \] 

The variance of the wavelet coefficients at a given scale $\tau_j = 2^{j-1} \bar{\Delta}; \nu_X^2 (\tau_j) = var(W_{j,t})$  gives the contribution to the total variance of the time series due to scale $\tau_j$: $var(X_t) = \sum_{j=1}^\infty \nu^2_X(\tau_j)$. Characteristic scales are indicated by peaks or changes of behavior in $\log (\nu^2_X) \, \mathrm{vs.} \, \log(\tau_j)$. Although wavelets traditionally require regularly sampled time series, Slepian wavelets (\cite[Mondal \& Percival 2011]{mondal11}) work with irregular and gappy time series. \cite[Graham et al. (2014)]{graham14} demonstrated that quasars and stars exhibit different Slepian wavelet variance profiles with quasars showing more variation on observed timescales greater than $\sim150$ days. They also found evidence for an intrinsic timescale deviation from a CAR(1) model that anti-correlated with luminosity. Further analysis is under way to understand the origin of this result and how it relates to characteristic timescales detected by other statistical analyses, e.g., those associated with CARMA models.

\subsection{Further options}

Many of the statistical descriptors commonly used to characterize astronomical time series are discriminative -- learning the boundaries between classes -- rather than generative -- modelling the individual distributions of classes. Though the former may be easier to compute and be more suitable for classification purposes, the latter offer more potential insight into the physical processes underpinning the time series. We are exploring a number of state of the art generative algorithms applied to CRTS quasar light curves:

\begin{itemize}

\item {\it Echo state networks} are a recurrent neural network that aim to capture the latent regime (physical processes) underlying a time series (see Giannotis et al., these proceedings)

\item {\it Empirical mode decomposition} (also known as the Hilbert-Huang transform) decomposes a time series into a set of data-defined intrinsic mode functions (IMFs) and a residual trend. It is then possible to determine the deterministic and stochastic components contribution to the time series, e.g., by looking at the mutual information between successive IMFs.

\item {\it Continuous time autoregressive fractionally integrated moving average (CARFIMA)} models are the superset to which CARMA models belong (CARMA(p,q) = CARFIMA(p, 0, q)) (see Feigelson, these proceedings).

\item {\it Symbolic regression} aims to find the best fitting analytical function to describe a data set. \cite[Graham et al. (2013)]{graham13} demonstrated its application to a number of astronomical data sets to recover known relationships such as the fundamental plane of elliptical galaxies and also as a binary classifier. \cite[Krone-Martins et al. (2014)]{krone14} have also used it to provide a functional form for photometric redshifts.

\end{itemize}

\section{Extreme variability} 

Given a sufficiently large sample, it is possible to parameterize the distribution of any characterizing variability features and identify extreme objects, i.e., those that belong in the tails or are outliers (note that if the tail is heavy then a specific model for it must be employed, e.g., Cauchy, otherwise it is impossible to determine significance for a generic heavy tail and outliers are possible without a heavy tail \cite[(Klebanov 2016)]{klebanov16}). This then allows an investigation of whether there is a clear transition from regular to extreme variability in quasars and what extreme variability reveals about accretion physics and the AGN/galaxy connection.

We are employing three different techniques to detect extreme sources:

\begin{itemize}
\item for characterizing features with unimodal symmetric distributions, the covariance is estimated via minimum covariance determinant and then the Mahalanobis distance is used to identify sources of interest;

\item for higher dimensional behaviours, the TSNE is calculated for both sets of features and time series themselves (following regularization with a Gaussian process) and then distinct clusters and non-members identified;

\item and finally {\it topological} outliers are found from an inferred persistent homology using the minimal spanning tree construct of the sample.

\end{itemize}

From these we are studying three distinct phenomena.

\begin{figure}[b]
\begin{center}
\includegraphics[width=5.3in]{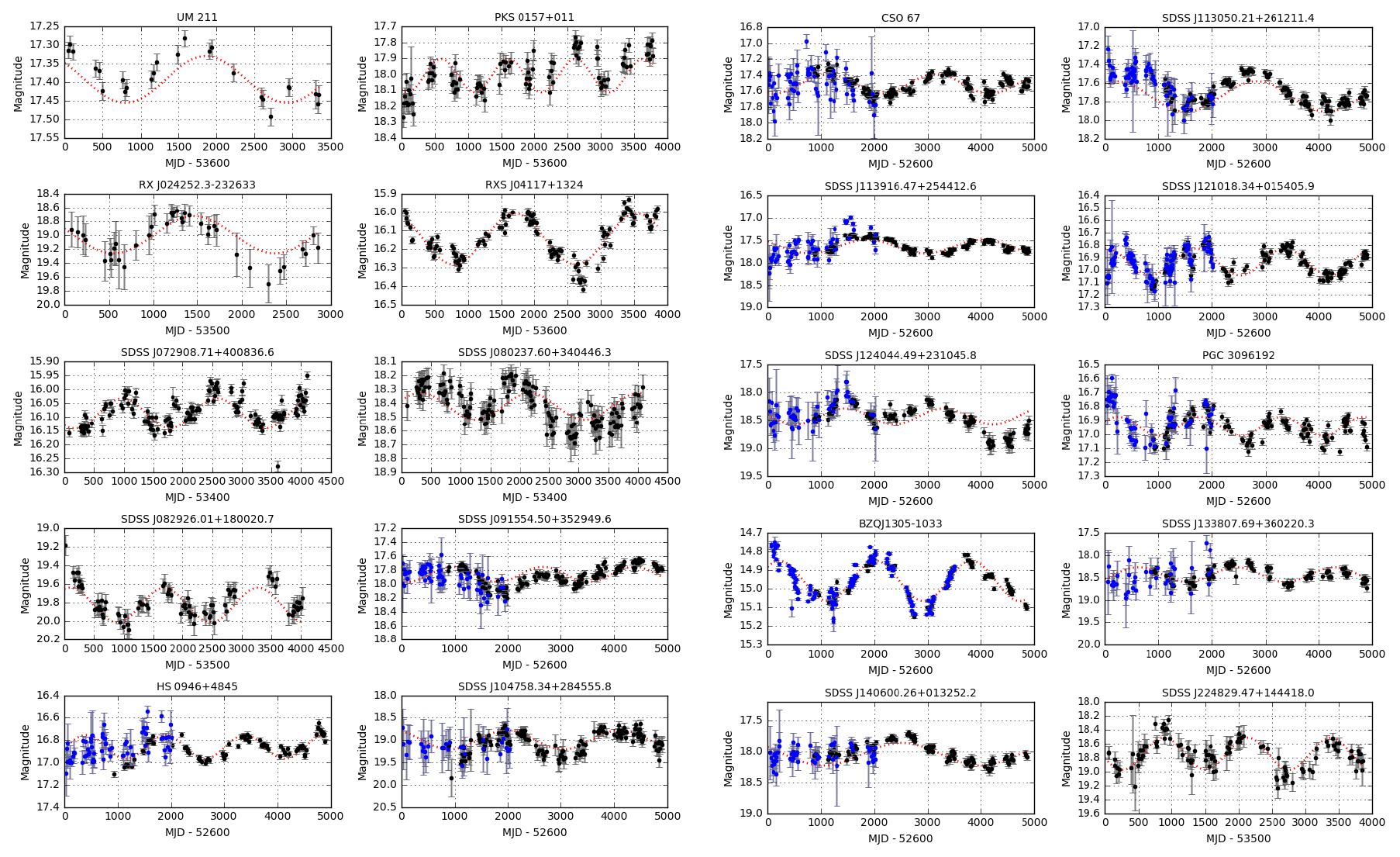} 
\caption{A sample of supermassive black hole binary candidates from \cite[Graham et al. (2015b)]{graham15b}. The blue points indicate data from the LINEAR survey and the black from CRTS. The dotted red line is the best fit sinusoid at the detected period for illustration purposes only.}
\label{fig2}
\end{center}
\end{figure}

\subsection{Supermassive black hole binaries}

Supermassive black hole (SMBH) binaries are an expected consequence of hierarchical galaxy formation models but the closest pairs (sub-parsec separation) are not detectable by direct imaging. Instead, recent searches have concentrated on identifying temporal changes in spectroscopy and photometry that are consistent with binary motion. \cite[Graham et al. (2015a, b)]{graham15a, graham15b} reported a sample of 111 quasars with statistically significant periodicity in their CRTS light curves (see Fig.~\ref{fig2}). Assuming that the detected period is the binary rotation period, these sources are all at sub-parsec separations and so in the final stages of merging. Although the underlying physical mechanism(s) for the periodicity seen is not easily identifiable -- jet precession, warped accretion disk, relativistic Doppler boosting, or periodic accretion -- these all imply a SMBH binary system, i.e., there is no physical mechanism attributable to a single SMBH system that works.

We are now engaged in a spectroscopic monitoring campaign of the periodic candidates to look for any associated spectroscopic variability. At such small separations ($\sim 0.01$pc), it is unlikely that emission lines from the two broad line regions associated with the SMBHs would be resolvable -- their separation is likely to be smaller than their velocity widths. However, the effects of periodic accretion flows onto the binary (\cite[Artymowicz \& Lubow 1996; Hayasaki, Mineshige, \& Ho 2008]){artymowicz96,hayasaki08} or radiative transfer effects, capable of driving winds and outflows in the circumbinary region (\cite[Nguyen \& Bogdanovich 2016]{nguyen16}), may be directly detectable.

Quasar variability is not a white noise process, i.e., its power spectrum is not flat ($P(f) \neq const$), and red noise components (such as DRW and CARMA) can introduce false periodicities or boost particular frequency ranges in a variogram. It is important this is treated properly in any period searching algorithm (\cite[Vaughan et al. 2016]{vaughan16}), either by explicitly having a null hypothesis assuming just a red noise background or through appropriate simulations. We are extending our searches for periodicity in quasars to include this, e.g., using periodic red noise kernels with Gaussian process regression (Dennis et al., in prep.) or through appropriate population simulations, as well as looking for periodicity in combined data sets, e.g., LINEAR + CRTS or PTF + CRTS, to potentially increase the temporal baseline and get additional photometry.

Close SMBH binaries are predicted to be the main source of nanofrequency gravitational waves. If the sample is taken as representative, its restframe period distribution implies a population of unequal mass SMBH binaries with a typical mass ratio as low as $q \sim 0.01$ (\cite[Charisi et al. 2016]{charisi16}). This also suggests that circumbinary gas is present at small orbital radii and is being perturbed by the black holes. These sources are potentially resolvable as continuous gravitational wave sources using the next generation pulsar timing arrays.

\begin{figure}[h]
\begin{center}
\includegraphics[width=5.1in]{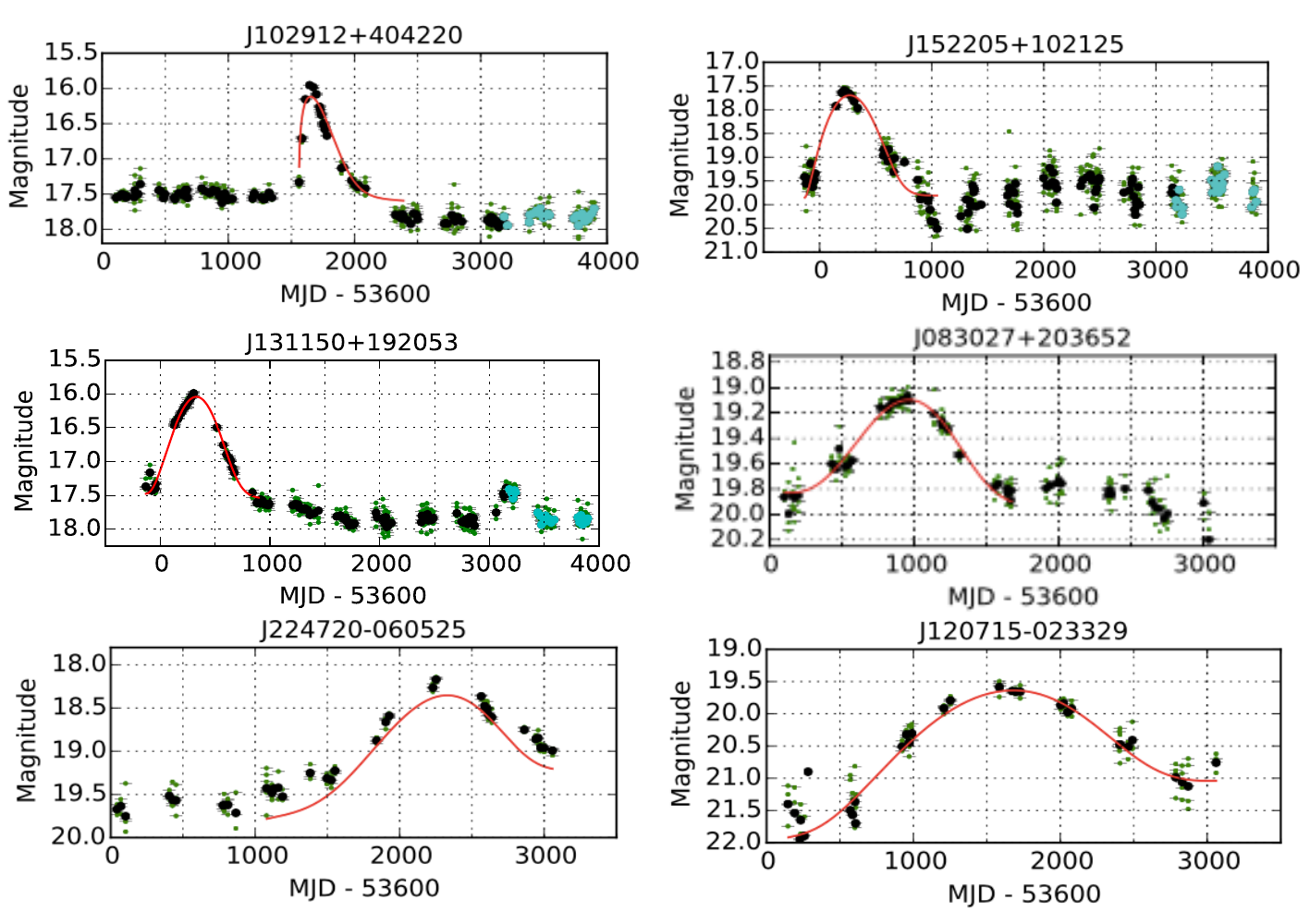} 
\caption{A sample of AGNs with major flares from \cite[Graham et al. (2017)]{graham17}. The red line is the best fit Weibull distribution to the data.}
\label{fig3}
\end{center}
\end{figure}

\subsection{Major flares}

Significant optical/UV outbursts have been reported in quiescent galaxies, consistent with superluminous supernovae (SLSNe) or candidate tidal disruption events (TDEs) (\cite[Gezari et al. 2012; Chornock et al. 2014; Komossa et al. 2015]{gezari12, chornock14, komossa15}). Those associated with active galaxies, though, are much rarer (\cite[Meusinger et al. 2010; Drake et al. 2011]{meusinger10, drake11}) as a single significant event can be hard to separate from general AGN variability. Such events, however, can provide insight into the structure and mechanics of the accretion disk and nuclear region.

We have identified 51 sources from over 900,000 known quasars and high probability quasar candidates which show a single major flaring event from a quiescent state (see Fig.~\ref{fig3} for a sample; \cite[Graham et al. 2017]{graham17}). The events typically last 900 days (in the observed frame) and have a median peak amplitude of $\Delta m = 1.25$ mag. We characterize the flare profile with a Weibull distribution and find that there are three distinct shapes: symmetric, fast rise exponential decay, and slow rise fast decay. This suggests that there is no single physical mechanism for all the flares. 

\cite[Lawrence et al. (2016)]{lawrence16} reported a search for large amplitude ($\Delta m >$ 1.5 mag) nuclear changes in faint extragalactic objects over a 10-year baseline from Pan-STARRS and SDSS data. 43 AGN were identified showing smooth order of magnitude outbursts over several years and large amplitude microlensing by stars in foreground galaxies is favored as the most likely explanation. We simulated 100,000 single-point single lens models with the same data priors as the CRTS data set and found that the distribution of Weibull characterizations for these was different than for most of our flare sample. We have also determined best-fit MCMC single-point single lens models to our flares and find that nine are well-described by such a model.

Other possible explanations for the flare events are superluminous IIn supernovae and slow TDEs, associated with a spinning supermassive black hole and a debris stream that is only self-interacting after several windings (\cite[Guillochon \& Ramirez-Ruiz (2015)]{guillochon16}). The majority of our events can thus be attributed to explosive stellar-related activity in the accretion disk. This suggests a class of extreme phenomena distinct from general quasar variability.

\begin{figure}[t]
\begin{center}
\includegraphics[width=5.1in]{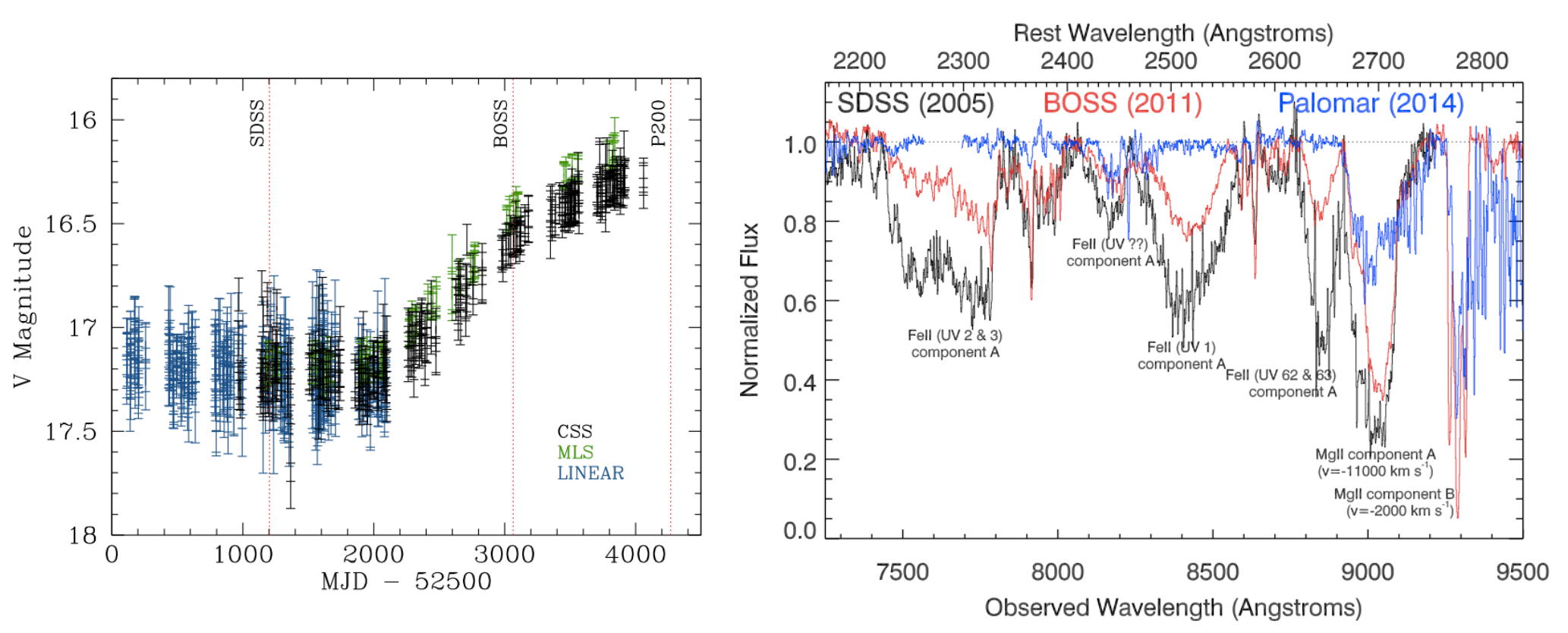} 
\includegraphics[width=5.2in]{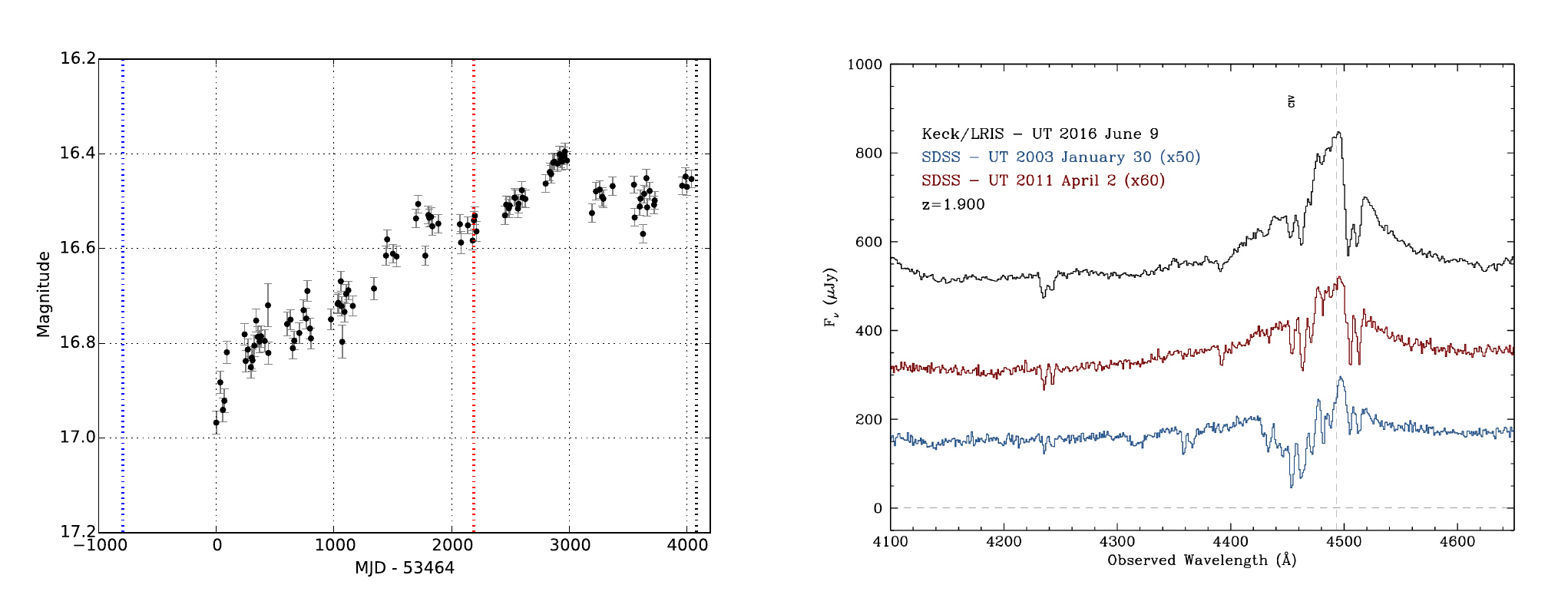} 
\caption{Two examples of quasars with steady rises in their light curves and the associated spectral variability. The top source is a BALQSO with varying absorption (from \cite[Stern et al. 2016]{stern16}); the lower source has an emerging CIV component. The colored lines indicate the different epochs at which spectra were obtained.}
\label{fig4}
\end{center}
\end{figure}

\subsection{Changing-look quasars}

A handful of objects -- so-called changing look quasars (CLQs) -- have been reported showing slow but consistent photometric variability ($\Delta m > 1$ mag) over several years coupled with spectral variability (\cite[LaMassa et al. 2015, Ruan et al. 2016, MacLeod et al. 2016, Runnoe et al. 2016]{lamassa15, ruan16, macleod16, runnoe16}). Their optical spectra show emerging or disappearing broad emission line components, typically H$\beta$. This is consistent with a change of type (Type 1 - Type 1.2/1.5 to 1.8/1.9 - Type 2 or vice versa) and may be associated with a large change of obscuration or accretion rate. 

We have identified a set of CRTS AGN which show a steady photometric rise/decline of $\Delta m \sim 1$ mag over several years between states of lower or higher activity. We are now obtaining spectra for these to look for spectroscopic variability associated with these changes. Fig.~\ref{fig4} shows examples of differing behaviors seen. The first source is a FeLoBAL with time-varying absorption trough depths (\cite[Stern et al. 2016]{stern16}) which suggests changes in photoionization equilibrium. The second source is more consistent with a standard CLQ with an emerging CIV component. Large accretion rate changes caused by cold chaotic accretion models or metallicity could be responsible for this type of activity. Further X-ray observations of such sources can isolate nuclear emission and probe the total absorption column thus helping to distinguish between explanations. 

\section{Quasar selection}

Quasars have a distinctive spectral energy distribution with a strong redshift dependency which has made color the most common selection criterion to date. The recent availability of mid-IR all-sky photometry has seen {\it WISE} colors replace SDSS and NIR-based techniques as the most highly rated method. Variability offers an alternative selection criterion that is independent of SED features. 
\cite[Graham et al. (2014)]{graham14} showed that $\sim10-20$\% of the quasar population would only be detected by variability and $\sim5 - 25$\% only by their {\it WISE} colors. Combining SED-based features (e.g., colors) with variability-based ones for selection purposes will therefore produce much more complete catalogs of sources. 

\begin{figure}[t]
\begin{center}
\includegraphics[width=5.0in]{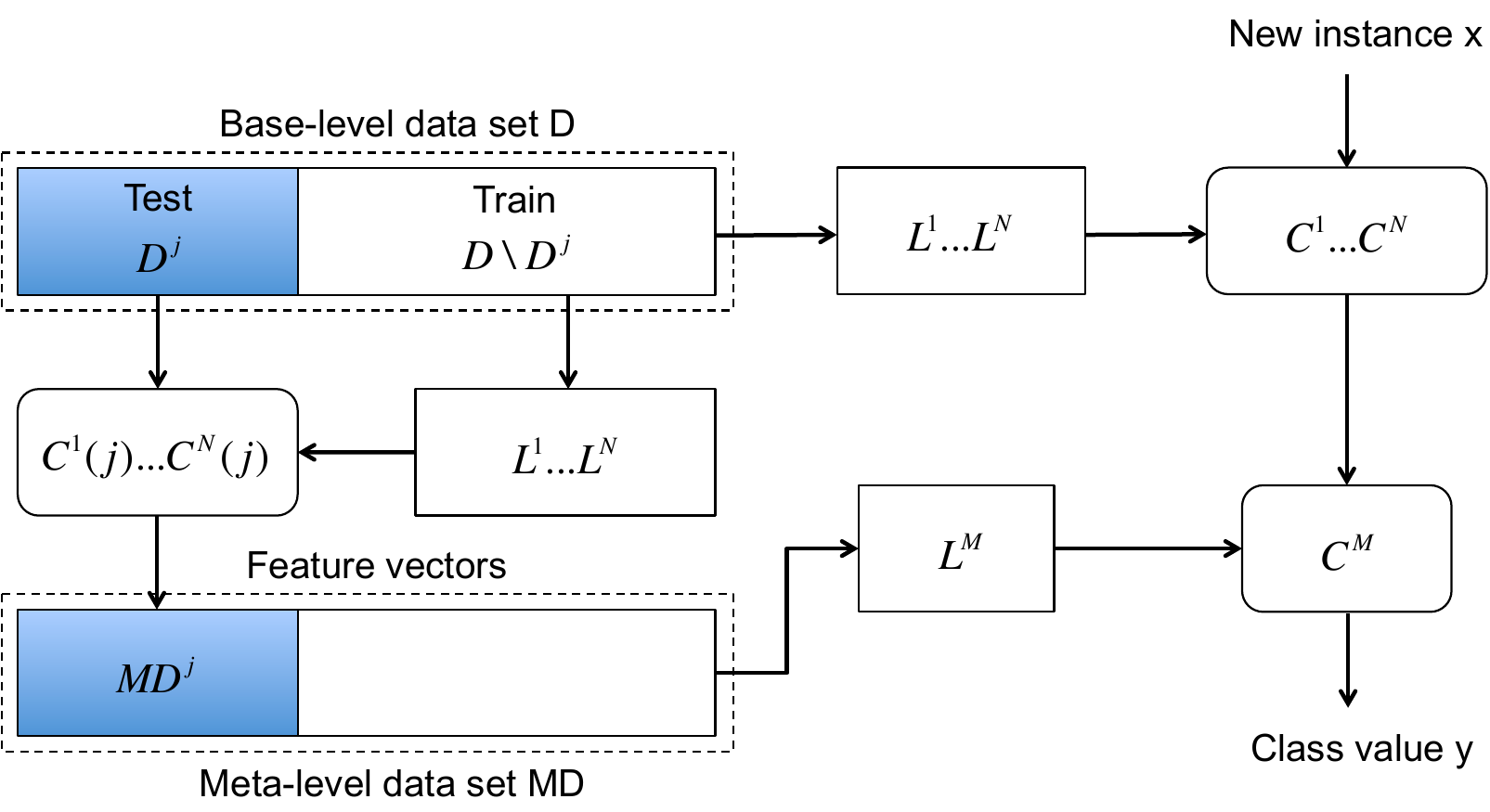} 
\caption{Stacked architecture for quasar selection. The base level data set consists of variability measures and colors for a set of known quasars. For each $j^{th}$ fold, the set of $N$ learning algorithms, $L$, is applied to the training part $D/D^j$ and $N$ induced classifiers, $C^N$, to the test part $D^j$. The concatenated predictions together with the original class prediction then form a new set of meta-level data, $MD$. A learning algorithm $L^M$ is trained on this to induce the meta-level classifier $C^M$ and the $L^N$ algorithms are trained on the full data set $D$. When a new instance arrives, the concatenated predictions of the $N$ base-level classifiers $C^N$ form the meta-level vector to which the meta-level classifier $C^M$ assigns a class value.}
\label{fig5}
\end{center}
\end{figure}

\begin{figure}[h]
\begin{center}
\includegraphics[width=2.6in]{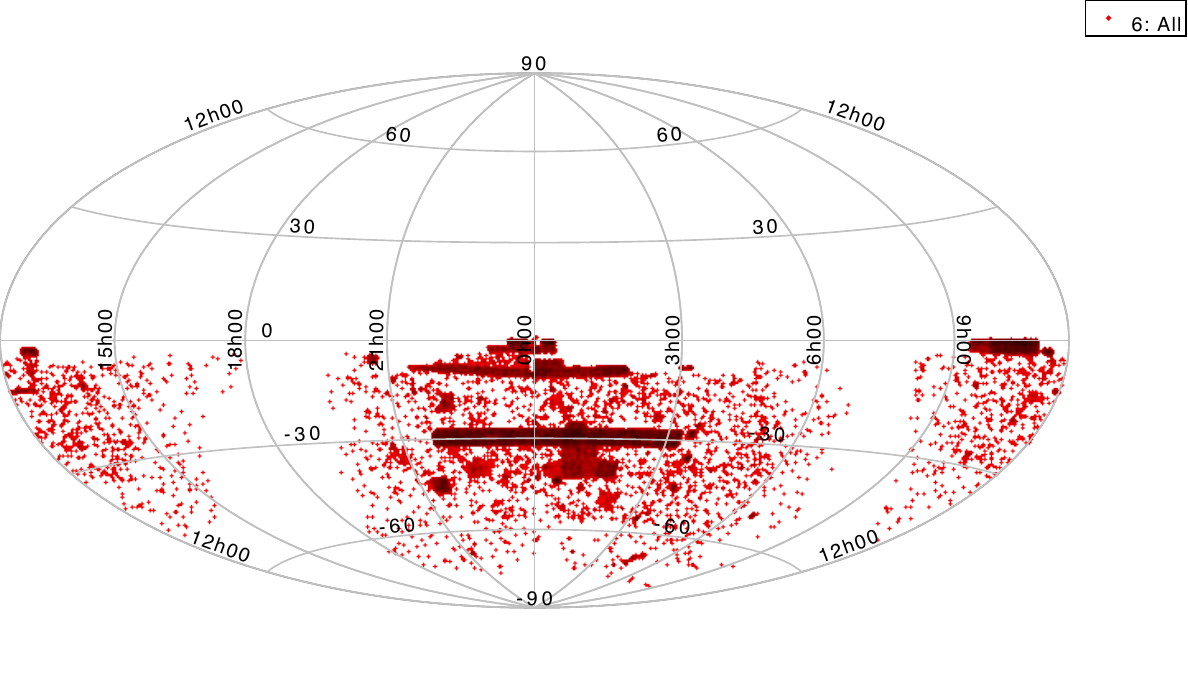} 
\includegraphics[width=2.4in]{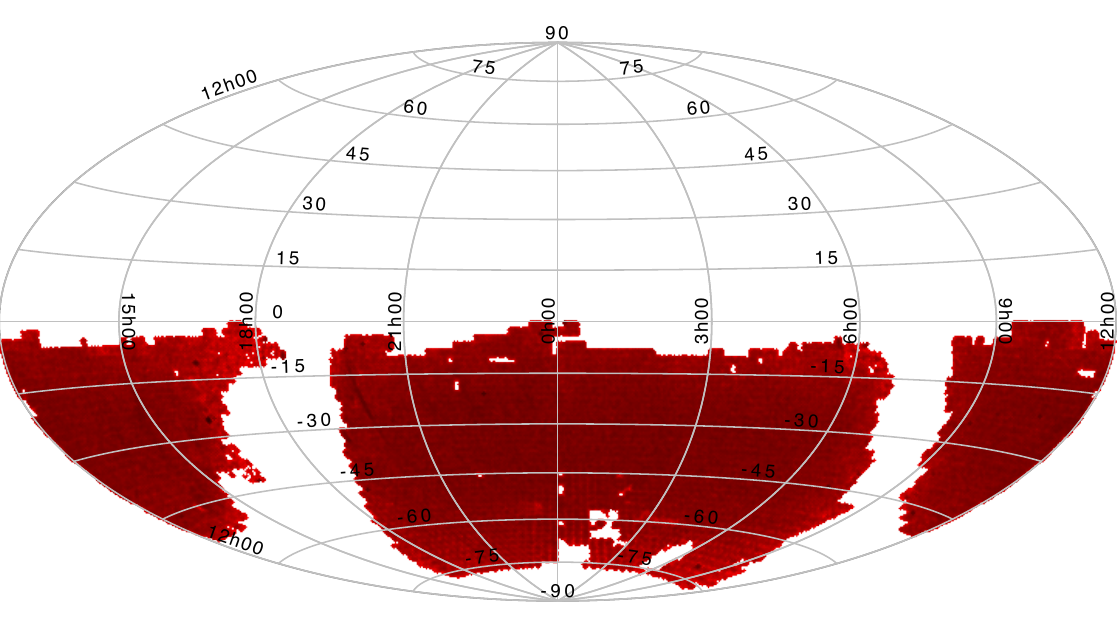} 
\caption{The distribution of known quasars within the SSS footprint in the southern hemisphere (left) and the distribution of high-probability candidates from the ensemble classifier applied to the SSS data set.}
\label{fig6}
\end{center}
\end{figure}

We are using a stacked ensemble classifier (see Fig.~\ref{fig5}) to combine {\it WISE} colors with variability features from structure functions, autoregressive models, and Slepian wavelet variance. The completeness and purity values of each individual method range from $65 - 90\%$  but for the ensemble classifier we find values of $\sim99\%$ for both from cross-validation tests using a set of 240,000 known quasars. Our eventual aim is to produce a catalog across the full 33,000 deg$^2$ coverage of CRTS but initially we are focusing on the southern sky coverage from SSS. Fig.~\ref{fig6}
shows the distributions of known quasars in the southern sky ($\sim 25,800$ within the SSS footprint) and of our $\sim 450,000$ high probability quasar candidates (to $V \sim 19.5$). We are also applying joint color-variability selection techniques to identify AGN in Kepler K2 fields.

\section{Conclusions}
Astroinformatics involves the application of machine learning and advanced statistical techniques to enable scientific discovery in large data sets. The detection of quasars in the time domain can be seen as a prime example. Large amounts of multiepoch sky coverage are required to provide populations temporally characterizable beyond simple measures of variability. However, within these data sets, 
it is possible to both identify new modes of behavior as well as discover unusual objects, 
providing new insights into physical models. These also inform on new methodologies to further increase population sizes extending past traditional detection thresholds. Finally, combining, exploring and mining the output of the upcoming generation of facilities -- LSST, eROSITA, SKA, and WFIRST -- will provide a multiwavelength understanding of quasar variability.





\begin{thebibliography}{}


\bibitem[Aretxaga et al. 1997]{aretxaga} {Aretxaga, I., Cid Fernandes, R., Terlevich R.,} 1997, \textit{MNRAS}, 286, 271
\bibitem[Artymowicz \& Lubow 1996]{artymowicz96} {Artymowicz, P., Lubow, S.H.,} 1996, \textit{ApJ}, 467, 77
\bibitem[Bauer et al. 2009]{bauer09} {Bauer A., et al.,} 2009, \textit{ApJ}, 696, 1241
\bibitem[Chornock et al. 2014]{chornock14} {Chornock R., et al.,} 2014, \textit{ApJ}, 780, 44
\bibitem[Cid Fernandes et al. 1997]{cidfernandes97} {Cid Fernandes, R., Terlevich, R., Aretxaga, I.,} 1997, \textit{MNRAS}, 289, 318
\bibitem[Drake et al. 2009]{drake09} {Drake, A.J., et al.}, 2009, \textit{ApJ}, 696, 870
\bibitem[Drake et al. 2011]{drake11} {Drake, A.J., et al.,} 2011, \textit{ApJ}, 735, 106
\bibitem[Drake et al. (2013)]{drake13} {Drake, A.J., et al.,} 2013, \textit{ApJ}, 763, 32
\bibitem[Gezari et al. 2012]{gezari12} {Gezari S., et al.,} 2012, \textit{Nature}, 485, 217
\bibitem[Graham et al. (2013)]{graham13}  {Graham, M.J., et al.,} 2013, \textit{MNRAS}, 
\bibitem[Graham et al. 2014]{graham14} {Graham, M.J., et al.,} 2014, \textit{MNRAS},
\bibitem[Graham et al. (2015)]{graham15a} {Graham, M.J., et al.,} 2015a, \textit{Nature},
\bibitem[Graham et al. 2015]{graham15b} {Graham, M.J., et al.,} 2015b, \textit{MNRAS},
\bibitem[Graham et al. 2017]{graham17} {Graham, M.J.}, 2017, \textit{MNRAS}, submitted
\bibitem[Guillochon \& Ramirez-Ruiz (2015)]{guillochon15} {Guillochon, J., Ramirez-Ruiz, E.,} 2015, \textit{ApJ}, 809, 166
\bibitem[Hawkins 1993]{hawkins93} {Hawkins, M.R.S.,} 1993, \textit{Nature}, 366, 242
\bibitem[Hawkins 2010]{hawkins10} {Hawkins, M.R.S.,} 2010, \textit{MNRAS}, 405, 1940
\bibitem[Hayasaki, Mineshige, \& Ho 2008]{hayasaki08} {Hayasaki, K., Mineshige, S., Ho, L.C.,} 2008, \textit{ApJ}, 682, 1134
\bibitem[Kasliwal et al. 2016]{kasliwal16} {Kasliwal, V.P., Vogeley. M.S., Richards, G.T.,} 2016, arXiv:1607.04299
\bibitem[Kawaguchi 1998]{kawaguchi} {Kawaguchi, T., et al.,} 1998, \textit{ApJ}, 504, 671
\bibitem[Kelly et al. 2009]{kelly09} {Kelly, B.C., Bechtold, J., Siemiginowska, A.,} 2009, \textit{ApJ}, 698, 895
\bibitem[Kelly et al. 2013]{kelly13} {Kelly, B.C., et al.,} 2013, \textit{ApJ}, 779, 187
\bibitem[Kelly et al. 2015]{kelly14} {Kelly, B.C., et al.,} 2014, \textit{ApJ}, 788, 33
\bibitem[Klebanov 2016]{klebanov16} {Klebanov, L.B.,} 2016, arXiv:1611.05410
\bibitem[Komossa et al. 2015]{komossa15} {Komossa., S., et al.,} 2015, \textit{A\&A}, 574, 121
\bibitem[Kozlowski 2016b]{kozlowski16b} {Kozlowski, S.,} 2016a, \textit{MNRAS}, 458, 2787
\bibitem[Kozlowski (2016a)]{kozlowski16b}  {Kozlowski, S.,} 2016b, arXiv:1611.08248
\bibitem[Krone-Martins et al. 2014]{krone14} {Krone-Martins, A., Ishida, E.E.O., de Souza, R.S.,} 2014, \textit{MNRAS}, 443, L34
\bibitem[Lamassa et al. 2015]{lamassa15} {LaMassa, S.M., et al.,} 2015, \textit{ApJ}, 800, 144
\bibitem[Lawrence et al. 2016]{lawrence16} {Lawrence A., et al.,} 2016, \textit{MNRAS}, 
\bibitem[MacLeod et al. 2010]{macleod10} {MacLeod, C., et al.,} 2010, \textit{AJ}, 721, 1014
\bibitem[MacLeod et al. 2016]{macleod16} {MacLeod, C., et al.,} 2016, \textit{ApJ}, 457, 389
\bibitem[Matthews \& Sandage (1963)]{3c48} {Matthews, T., Sandage, A.,} 1963, \textit{ApJ}, 46, 138
\bibitem[Meusinger et al. 2010]{meusinger10} {Meusinger, H., et al.,} 2010, \textit{A\&A}, 512, A1
\bibitem[Nguyen \& Bogdanovic (2016)]{nguyen16} {Nguyen, K. Bogdanovic, T.}, 2016, \textit{ApJ}, 828, 68
\bibitem[Mondal \& Percival 2011]{percival08} {Mondal, D., Percival, D.B.,} 2011, in \textit{Statistical Challenges in Modern Astronomy V}, eds. Ferguson, E.D., Babu, G.J., Springer, New York, pp. 403
\bibitem[Ruan et al. 2016]{ruan16} {Ruan, J.J., et al.,} 2016, \textit{ApJ}, 826, 188
\bibitem[Runnoe et al. 2016]{runnoe16} {Runnoe, J.C., et al.,} 2016, \textit{MNRAS}, 455, 1691
\bibitem[Schmidt et al. 2010]{schmidt10} {Schmidt, K.B., et al.,} 2010, \textit{ApJ}, 714, 1194
\bibitem[Schmidt et al. 2012]{schmidt12} {Schmidt, K.B., et al.,} 2012, \textit{ApJ}, 744, 147
\bibitem[Stern et al. 2016]{stern16} {Stern, D., et al.,} 2016, \textit{ApJ}, submitted
\bibitem[Torricelli et al. 2000]{torricelli00} {Torricelli-Ciamponi, G., et al.,} 2000, \textit{A\&A}, 358, 57
\bibitem[Ulrich et al. 1997]{ulrich} {Ulrich, M.-H., Maraschi, L., Urry, C.M.,} 1997, \textit{ARA\&A}, 35, 445
\bibitem[Vanden Berk et al. 2004]{vandenberk04} {Vanden Berk, D.E., et al.,} 2004, \textit{ApJ}, 601, 692
\bibitem[van der Maarten \& Hinton (2008)]{maarten08}  {van der Maarten, L.J.P., Hinton, G.}, 2008, \textit{Jour. Machine Learning Research}, 9, 2579
\bibitem[Vaughan \etal (2016)]{vaughan16} {Vaughan, S., et al.,} 2016, MNRAS, 461, 3145

\end{thebibliography}
\end{document}